\documentstyle[preprint,aps,psfig,eqsecnum]{revtex}
\begin{document}

\draft
\tightenlines

\preprint{\vbox{\hbox{U. of Iowa preprint 97-2502; CERN 97-252}}}

\title{A Check of a $D=4$ Field-Theoretical Calculation 
Using the High-Temperature Expansion 
for Dyson's Hierarchical Model }

\author{J. J. Godina\\
{\it Dep. de Fis. , CINVESTAV-IPN , Ap. Post. 14-740, Mexico, D.F. 07000\\
and\\
Dpt. of Physics and Astr., Univ. of Iowa, Iowa City, Iowa 52246, USA}}
\author{Y. Meurice\cite{byline}\\ 
{\it CERN, 1211 Geneva 23, Switzerland\\ 
and\\ 
Dpt. of Physics and Astr., Univ. of Iowa, Iowa City, Iowa 52246, USA}}
\author{S. Niermann\\
{\it Dpt. of Physics and Astr., Univ. of Iowa, Iowa City, Iowa 52246, USA}}

\maketitle
\begin{abstract}
We calculate the high-temperature expansion of the 2-point function
up to order 800 in $\beta$.
We show that estimations of  the 
critical 
exponent $\gamma $ based on asymptotic analysis 
are not very accurate in presence of confluent logarithmic singularities.
Using a direct comparison 
between the actual series and the series obtained from
a parametrization of the form $(\beta_c -\beta)^{-\gamma}
(Ln(\beta_c -\beta))^p +r)$, we show that the errors
are minimized for
$\gamma =0.9997$ and $p=0.3351$,
in very good agreement with field-theoretical calculations.
We briefly discuss the related questions of triviality and
hyperscaling.  
\end{abstract}

\newpage
\section{Introduction}
\label{sec:intro}

The dimension four plays a doubly important role in physics.
First, it is the dimension of space-time which is relevant for 
a relativistic
description of a large class of phenomena, from
electricity and magnetism 
to scattering  processes at the highest experimentally
accessible energies. Second, it is the upper critical
dimension for scalar field theory. If one analytically continues
the renormalization group equations\cite{wilson} 
(usually derived within some approximation) to non-integer dimensions,
it appears that when the dimension tends to four from below, the non-trivial
fixed point merges with the Gaussian one. This 
justifies the $\epsilon$-expansion.

It is thus commonly accepted that in four dimensions, the critical exponents
are the trivial ones (i.e. those obtained from mean field). 
Unfortunately, 
it often difficult to find  clear evidence for or against 
trivial exponents, for 
instance, from high-temperature (HT) series\cite{baker,sykes} 
or a finite volume calculation\cite{num}.
The root of the problem is the existence of a marginal 
direction which makes the approach to the fixed point more intricated 
than in three dimensions. 
The corrections to the power laws can in principle be obtained 
from the Callan-Symanzik equations, provided we know the exact form of the 
various functions (beta, gamma, ... ) entering into them.
Using the lowest order in perturbation theory,
Brezin, Le Guillou, and Zinn-Justin\cite{brezin} 
found that the trivial power divergences get multiplied by 
rational powers of $Ln(\beta_c -\beta)$. 
It is important to check this
result with methods independent of perturbation 
theory. In particular, it is conceivable that there exist non-trivial
fixed points which cannot be revealed by perturbation theory.

The technical challenge which appears in any kind of calculation 
is to distinguish between a small change (with respect to 
the trivial value) in the critical exponent 
and a slowly varying (compared to the trivial singularity) 
multiplicative change. This difficulty appears clearly 
in the asymptotic analysis of the high-temperature expansion
of the susceptibility, where
the leading term of the extrapolated slope defined in Eq. (\ref{eq:shat})
($\gamma -1$) can be small compared to corrections proportional
to the inverse of the logarithm of the order, unless one can 
reach an astronomically large order. 

Another interesting feature of the field-theoretical
method is the so-called hyperscaling relation among 
the power singularities of the 2- and 4-point 
(subtracted) Green's functions at 
zero momentum. In three dimensions, 
the violations of hyperscaling\cite{dis}
are hard to resolve by high-temperature calculation.
This is
still a controversial\cite{baker2} topic.
In four dimensions,
conflicting\cite{baker,sykes} conclusions
were drawn from the high-temperature series.

The confirmation of the field-theoretical results would 
require that an unbiased estimate of the main
power singularity {\it and} the power of the logarithmic correction
come close to their predicted values, with errors compatible
with (small) higher-order corrections. We propose here to test
the field-theoretical results using an expansion in the kinetic 
term (also called high-temperature expansion), 
in a model which is obviously non-trivial in three dimensions,
but where calculations are easier than in nearest-neighbor lattice models. 

The hierarchical model\cite{dyson} is a non-trivial 
approximation of models with
short range interactions, 
which is well-studied\cite{sinai,epsi}, and 
for which we can calculate the high-temperature
expansion\cite{high} 
to a very large order. The recursion relation which
summarizes the renormalization group transformation of
this model is closely related to the approximate
recursion formula discussed by Wilson\cite{wilson}.
The qualitative and quantitative aspects of this relationship
are discussed in Ref. \cite{fam}. 

In recent publications\cite{prl,osc}, we reported results concerning 
the high-temperature expansion of Dyson's hierarchical model
in three dimensions. We calculated the 
high-temperature expansion of the magnetic susceptibility
up to order 800 with Ising and 
Landau-Ginzburg
measures. 
This allowed us to obtain a value\cite{osc} of the 
critical exponent $\gamma$ of 1.300 in $D=3$, with estimated 
errors of order 0.002. This result is consistent with the results 
obtained with the $\epsilon$-expansion\cite{sinai,epsi}.

We found clear evidence for oscillations in the quantity, called the
 extrapolated slope\cite{nickel} (see section below), used
to estimate the critical exponent $\gamma $. 
When using a log scale for the order in the
high-temperature expansion, these oscillations become regularly 
spaced.
Our interpretation of the data was consistent with the hypothesis
that 
the eigenvalues of the linearized renormalization group
transformation are real, but that 
the constants
appearing in the conventional parametrization of the 
magnetic susceptibility should be replaced 
by functions of $\beta_c -\beta$ invariant
under the rescaling of $\beta_c -\beta$ by $\lambda _1$, 
the largest eigenvalue of 
the linearized renormalization group
transformation.
This possibility has been mentioned in the past
by K. Wilson\cite{wilson} 
and developed systematically by Niemeijer and 
van Leeuwen\cite{nie}.
Our analysis provided good evidence that the oscillations
appear with a universal frequency 
in good agreement with theoretical expectations,
but with a measure-dependent 
phase and amplitude. 

Subsequently, more efficient methods of calculation, based on finite
dimensional projections of the Fourier transform of the recursion 
formula, were developed. As explained in detail in 
Ref. \cite{finite}, the effects of such
truncations can be controlled with a precision which is better
than exponential when the dimension of the truncated space increases.

\begin{figure}
\vskip20pt
\centerline{\psfig{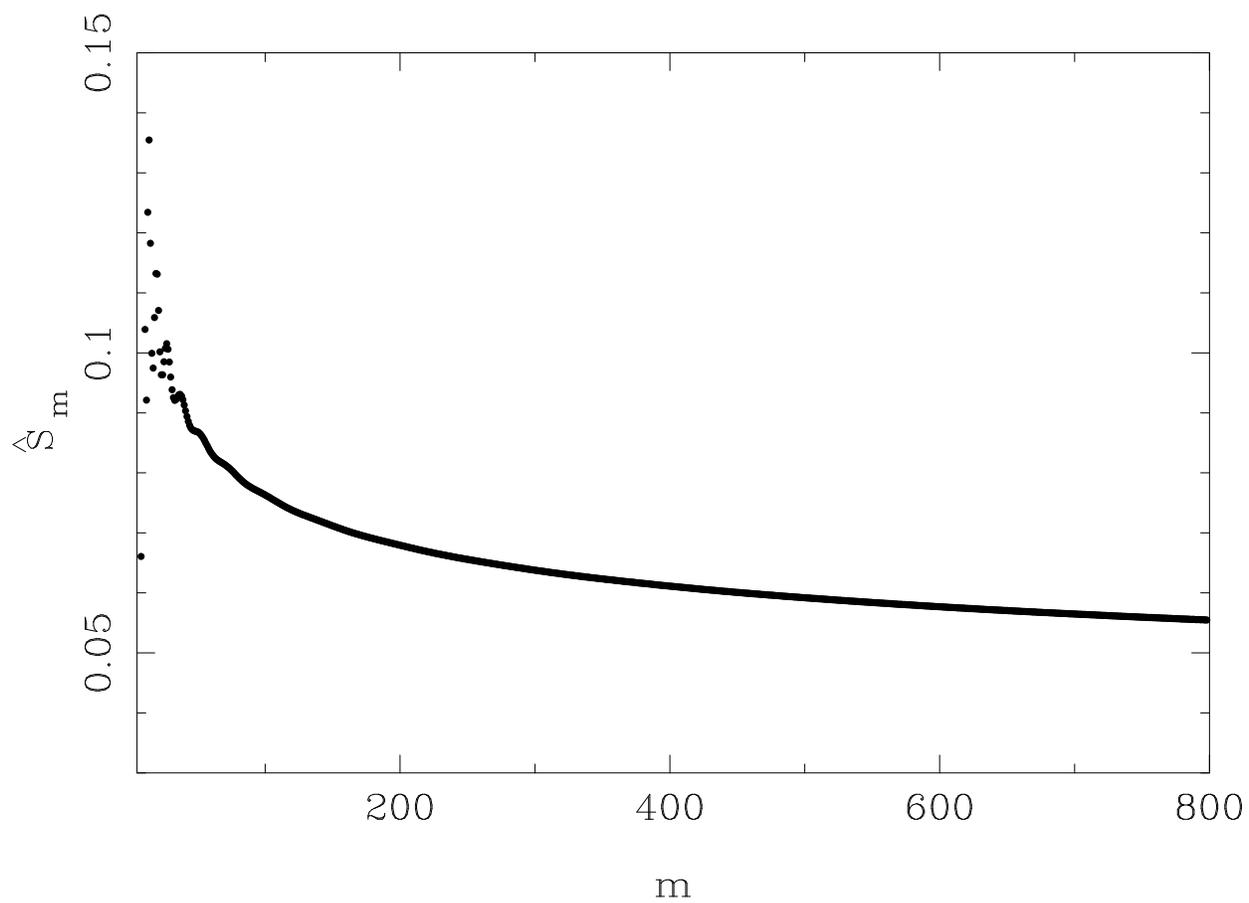}}
\vskip50pt
\caption{The extrapolated slope $\widehat{S}_m$ versus the order $m$ 
in the HT
expansion.}
\label{shat}
\end{figure}

In this paper, we study the high-temperature expansion of 
Dyson's hierarchical
model in dimension 4. 
For the sake of completeness, we briefly review
the method of calculation in section \ref{sec:calc}. 
The conventional methods\cite{nickel,gaunt} used to estimate the critical
temperature and a critical exponent from a high-temperature series
are reviewed in section \ref{sec:lim}. We show that in the presence of 
logarithmic corrections to the scaling laws, the asymptotic behavior of the
corrections is modified. The extrapolated ratio defined in 
Eq. (\ref{eq:rhat}) provides an estimate of the critical temperature
with corrections of order $m^{-1}\times (Ln(m))^{-2}$, where $m$ is
the order in the high-temperature expansion. In the following, we continue 
to use
the notation $m$ with the same meaning.
Using 
the  expansion of the susceptibility up to order 800, we 
obtained a value 
of the critical temperature 
which agreed with the high-precision determination of Ref. \cite{finite}
with errors of less than one part in 10,000. On the other hand, the 
extrapolated slope defined in Eq. (\ref{shat}) estimates the critical
exponent minus one with corrections which are only suppressed by
$(Ln(m))^{-1}$. If this weak suppression is not recognized, one 
may conclude that the  critical exponent $\gamma $ takes a value larger
than the trivial one. More generally, asymptotic analysis is not 
adequate to distinguish between a value of $\gamma$ close to 1 and a 
correction to the scaling laws which is less singular than a power.

In section \ref{sec:estgam}, we analyze 
the high-temperature expansion of the susceptibility
without relying on the asymptotic behavior of the coefficients.
We use $h(m)\equiv 
(r_m\beta_c-1)m$, a function which represents the difference 
between the ratio of successive coefficients $r_m$ and its asymptotic value
$\beta _c ^{-1}$. The function $h(m)$ can be calculated {\it exactly}
using either the empirical series or the series corresponding to a given
assumption on the analytical form of the susceptibility. Taking the sum 
over a large range of $m$ of 
the square of the differences between these two values of $(h(m))^{-1}$, 
one can get an error 
function which indicates how good the analytical assumption is.
We found that the parametrization
\begin{equation}
\chi=(\beta _c -\beta )^{-\gamma} (A_0 (-\beta ^{-1}
Ln(1-{\beta \over \beta_c}))^{p}+A_1)
\label{eq:guess} \end{equation}
provides very good fits 
of the data for $\gamma \simeq 1$ and $p \simeq {1\over 3}$, which
is the field-theoretical\cite{brezin} result. In order to decide 
how accurate
the agreement is, we have considered fixed values 
of $\gamma $ in the vicinity
of 1 and equally spaced by $10^{-4}$ steps. For each of these 
values, we have 
determined the values of $p$ and $A_1/A_0$ which minimize the 
error function.
This error function behaves like a paraboloid near its minimum 
at
$\gamma =0.9997$ and $p=0.3351$,
in good agreement with the field-theoretical calculation. The errors 
on this estimate are mostly systematic. To get more accurate
results, one needs to replace the constant $A_1$ by a slowly varying 
function.

Another quantity which can be studied using the high-temperature expansion
is the dimensionless renormalized coupling constant\cite{parisi}, 
denoted $\lambda_4$ hereafter, obtained by
multiplying the connected four-point function at zero momentum by
the eighth ($D+4$ in general) power of the renormalized mass. 
For $D<4$, this quantity
is designed to have a finite and non-zero limit when 
$\beta \rightarrow \beta_c$. In the case $D=4$, we have checked with
good accuracy\cite{finite} that
$\lambda_4$ goes to zero
like $(Ln(\beta_c -\beta ))^{-1}$ for the model studied here.
The calculation of the HT coefficients of $\lambda_4$ involves the 
subtraction of the disconnected part and suffers the same type of
numerical problems as the direct calculation of $\lambda_4$, as discussed in
Ref. \cite{finite}. For this reason, we were only able to extract a 
series of 30 coefficients. The analysis of this series is consistent
with the fact that $\lambda_4$ goes to zero  when 
$\beta \rightarrow \beta_c$ (triviality), but 
it is not possible to distinguish a $(Ln(\beta_c -\beta ))^{-1}$ approach 
to zero from a $(\beta_c -\beta )^{1/2}$ approach, which would
be necessary to establish whether or not hyperscaling holds.
This question has been settled in Ref. \cite{finite}, and this section
illustrates the inconclusiveness of results obtained from short series. 

In conclusion, we have shown that by using sufficiently long series and
methods of analysis not relying on an asymptotic 
expansion, it is possible
to obtain very good agreement between calculations based on field
theory and those based on high-temperature 
expansion in the upper critical dimension.
We emphasize that the main interest of the high-temperature expansion
is to allow us to probe global features of the renormalization group flows
which cannot be approached using renormalized perturbation theory or an 
analysis of the linearized behavior near the fixed point. An example 
of such 
a global feature is the existence of log-periodic 
oscillations\cite{prl,osc}, 
which play an important role in $D=3$, but have an almost negligible 
effect in $D=4$, as shown in section \ref{sec:lim}. Another example of a
global feature could be the existence of a non-trivial fixed point. 
The good agreement found in section \ref{sec:estgam} makes this possibility
very implausible for the model studied here.

\section{Calculations of the HT Coefficients}
\label{sec:calc}

The calculation of the high-temperature 
expansion of the unsubtracted $2k$-point functions of 
Dyson's hierarchical model
can be performed iteratively using the basic
recursion formula in its Fourier form \cite{high}. This method has been
discussed extensively in Refs. \cite{prl,osc}. For the sake
of being self-contained, we briefly explain the basic method of 
calculation. More details, justifications, and motivations can be found 
in Refs. \cite{num,prl,osc}.

The recursion formula for the 
rescaled Fourier transform $R_n(k)$ of the local 
measure for blocks of $2^n$ sites reads
\begin{equation}
R_{n+1}(k)=C_{n+1}\exp(-{1\over 2}\beta 
({c\over 4}s^2)^{n+1 }{{\partial ^2} \over 
{\partial k ^2}})(R_{n}({k\over s}))^2 \ , 
\label{eq:rec}\end{equation}
where $c$ is an adjustable parameter 
which takes the value $2^{1-{1\over D}}$,
in order to approximate $D$-dimensional models. In the following, 
we will only consider the case $D=4$, which means $c=\sqrt{2}$.
The rescaling operation
commutes with iterative integrations, and the 
rescaling factor $s$
can be fixed at our convenience. In order to obtain stabilized 
expressions in the high-temperature phase, we will take $s=\sqrt{2}$
in the following. 
We fix the normalization constant $C_n$ in such way
that $R_n(0)=1$. $R_n(k)$ then has a direct probabilistic
interpretation.
If we call $M_n$ the total field $\sum \phi _x$ inside
blocks of side $2^n$, and $<...>_n $
the average calculated without taking into account the interactions
among these blocks, we can write
\begin{equation}
R_n(k)=\sum_{q=0}^{\infty}{{(-ik)^{2q}}\over{2q!}} 
{<(M_n)^{2q}>_n\over{2^{qn}}} \ .
\label{gf} \end{equation}
We see that the Fourier transform of the local measure 
obtained after $n$ iterations generates the 
zero-momentum Green's functions calculated with 
$2^n$ sites.
All the calculations done here use an initial Ising measure, 
which means that $R_0(k)=cos(k)$. Since we are interested
in the leading singularity, this choice 
should play no role\cite{parisi} in the discussion.

The high-temperature expansion of the
zero-momentum Green's function
can be obtained
from an expansion of Eq. (\ref{eq:rec}) in powers of $\beta$. 
The most important sources of errors are
the round-off errors. After 
100 iterations, the relative errors 
on the $m$th coefficient\cite{osc} 
are of the order of $m\times 10^{-15}$.
With the choice $s=\sqrt{2}$, the coefficients reach a finite
value in the infinite volume limit. Actual computations are 
made at large but finite volume (i.\ e.\ at finite $n$).
The relative difference between the coefficients at finite and infinite $n$
goes to zero\cite{high} like $({c\over2})^n$.
For $D=4$, the choice $n=100$ means that $({c\over2})^n=2^{-50}$, which
is smaller than the numerical errors.

\begin{figure}
\centerline{\psfig{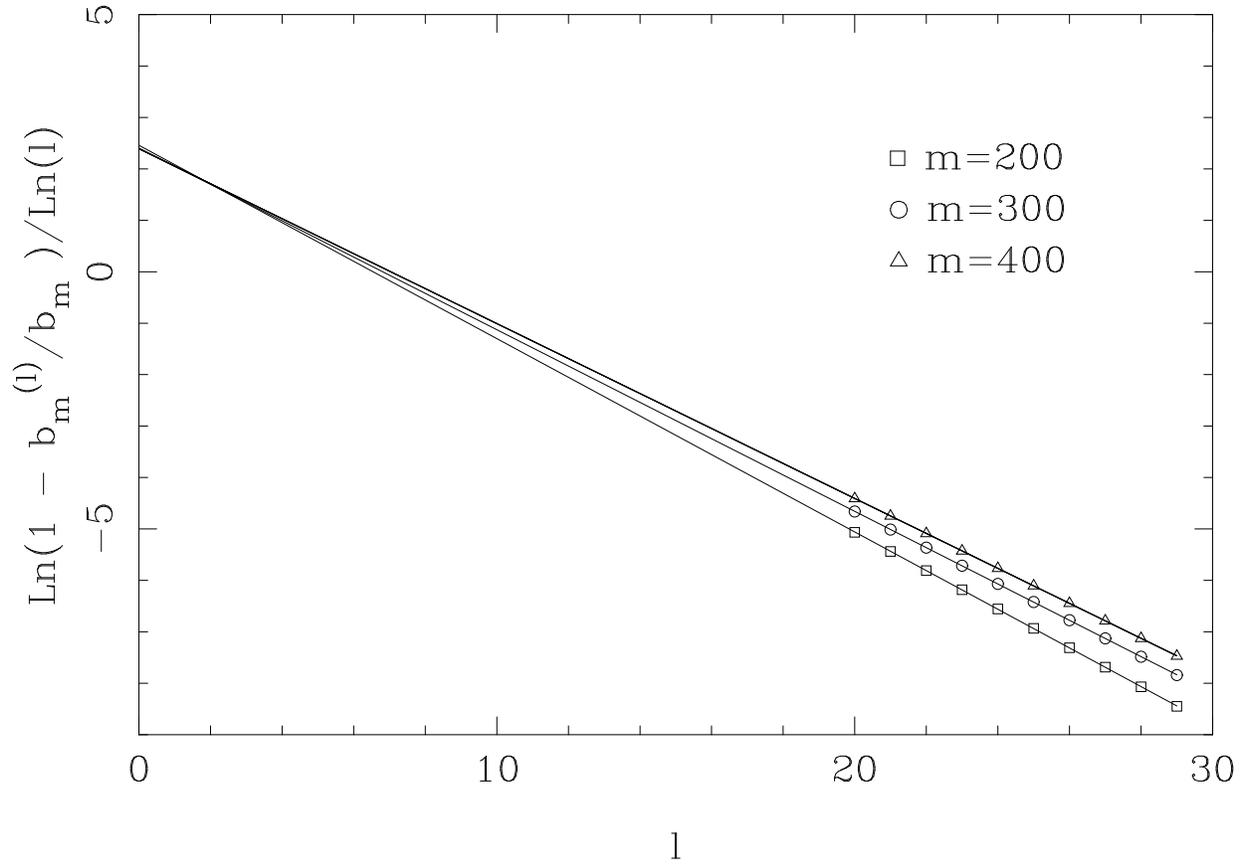}}
\vskip50pt
\caption{$l$-dependence of the high-temperature coefficients $b_m^{(l)}$
calculated in truncated spaces of dimension $l$.}
\vskip20pt
\label{ldep4}
\end{figure}

Such a calculation is in general time-consuming when one wants to calculate
more than 100 coefficients. It is, however, possible to save time
by using 
finite dimensional
approximations\cite{finite} of degree $l$ for the generating function:
\begin{equation}
R_n(k)=1+a_{n,1}k^2 +a_{n,2}k^4+..... +a_{n,l}k^{2l} \ ,
\label{eq:rn} \end{equation}
with $l$ much smaller than the required dimension $m+1$
necessary for an exact\cite{high} calculation.
After each iteration, non-zero coefficients of higher 
order ($a_{n+1,l+1}$ etc.\ ) are obtained, but 
set to zero in the next iteration.  
The $l$-dependence of the high-temperature
coefficients of the susceptibility is discussed  
in Ref. \cite{finite}.
If $b_{m}^{(l)}$ denotes the value 
of $b_{m}$ in a truncated space of dimension $l$, we found that
\begin{equation}
b_{m}^{(l)}=b_{m} (1-l^{-|s| l + q}) ,
\label{eq:ldep} \end{equation}
where $s$ and $i$ are, respectively, the slope and intercept of the
corresponding fitted line, as shown in Fig.\ \ref{ldep4}. 
The intercepts are approximately 2.3, while the slopes depend on $m$.
Eq.\ (\ref{eq:ldep}) represents suppressions which are better 
than exponential.
From this figure, we can check, for instance, that for $m=400$
(which is the maximal value used in section \ref{sec:estgam}), 
the extrapolated errors at $l=40$ are significantly lower than 
the numerical errors. Using extrapolation in $m$,
it was estimated in Ref. \cite{finite} that in the case $D=4$, $l=38$
was sufficient 
to calculate $b_{1000}$.

In summary, the following calculations will be 
performed with $l=50$ and $n=100$.
The above discussion shows that this choice guarantees that the systematic
errors are smaller than the numerical errors.

\section{The limitation of the Asymptotic Analysis in Presence of  
Confluent Logarithmic Singularities}
\label{sec:lim}

In this section, we study the singularities of the susceptibility
using its high-temperature
expansion
\begin{equation}
\chi(\beta)=\sum_{m=0}^{\infty} b_m \beta^m .
\end{equation}
We define $r_m=b_m/b_{m-1}$, the ratio of two
successive coefficients. 
When $D<4$, one expects\cite{parisi} that
\begin{equation}
\chi=(\beta _c -\beta )^{-\gamma } (A_0 + A_1 (\beta _c -\beta)^{
\Delta } +....)\ ,
\label{eq:convpar}  \end{equation}
and it is convenient to introduce 
quantities\cite{nickel} called the extrapolated ratio ($\widehat{R}_m$) 
and the  
extrapolated slope ($\widehat{S}_m$)
in order to estimate $\beta_c$ and $\gamma $.
These quantities are defined as
\begin{equation}
\widehat{R}_m = mr_m -(m-1)r_{m-1}\ ,
\label{eq:rhat}\end{equation}
and 
\begin{equation}
\widehat{S}_m  =  mS_m-(m-1)S_{m-1}\ ,
\label{eq:shat}\end{equation}
where 
\begin{equation}
S_m  = -m(m-1)(r_m - r_{m-1})/(mr_m -(m-1)r_{m-1})
\end{equation}
is called the normalized slope. When $A_0$
and $A_1$ are constant, one finds\cite{nickel} that
the $1/m$ corrections disappear:
\begin{equation}
\widehat{S}_m =\gamma -1 -B m^{-\Delta }+O(m^{-2}).
\label{eq:convdecay} \end{equation}

However, for the hierarchical model in $D=3$, 
large oscillations were observed\cite{prl} in $\widehat{S}_m$
and it was 
recognized\cite{prl,osc} that $A_0$
and $A_1$ should be considered as
functions of $\beta_c -\beta$ invariant
under the rescaling of $\beta_c -\beta$ by $\lambda _1$, 
the largest eigenvalue of 
the linearized renormalization group
transformation. The asymptotic analysis (when $m$ becomes large) 
of the extrapolated slope 
in this modified situation is given in section 3 of Ref. \cite{osc}.
It was found that $1/m$ corrections with rather 
large coefficients reappeared.
Nevertheless, it was possible to extract the critical exponent $\gamma $
with estimated errors of 0.2 percent.

\begin{figure}
\vskip20pt
\centerline{\psfig{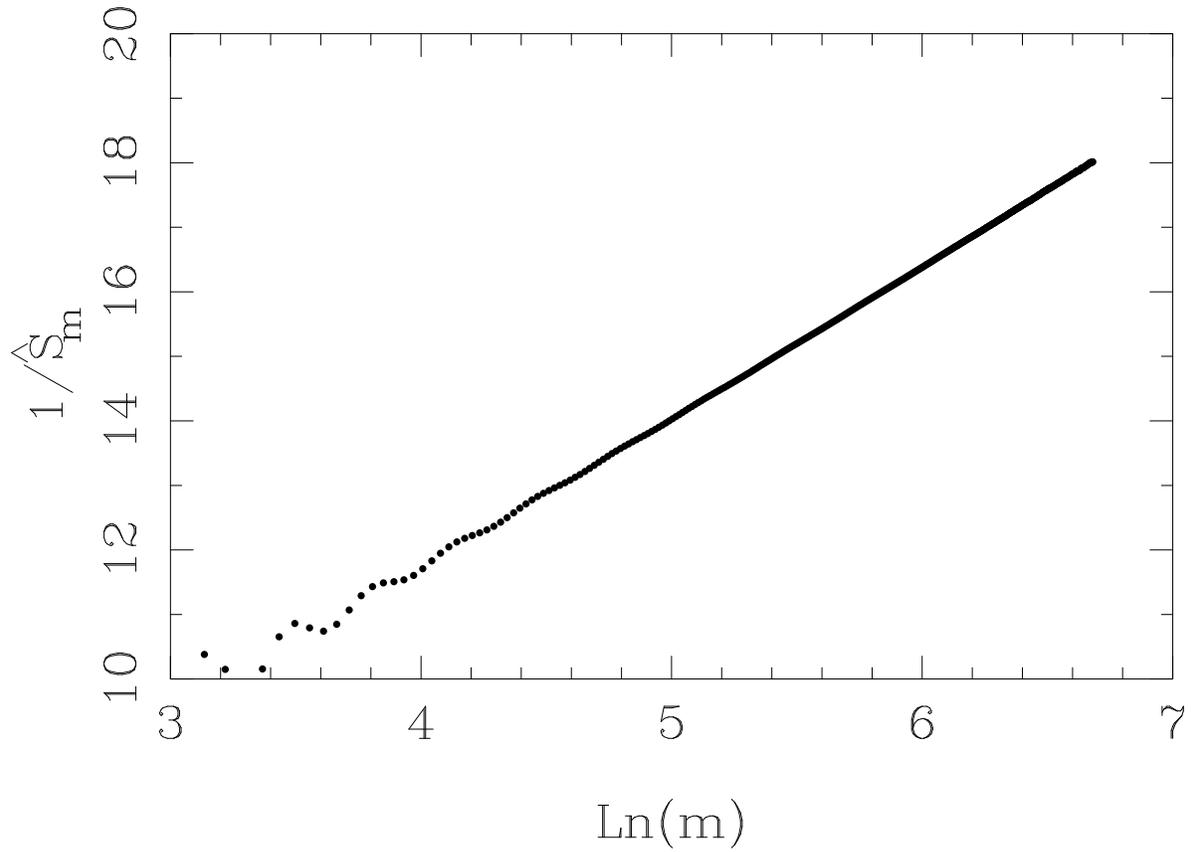}}
\vskip50pt
\caption{The inverse extrapolated slope $\widehat{S}_m$ versus 
the order $m$ in the HT. }
\label{invshat} 
\end{figure}

\begin{figure}
\vskip20pt
\centerline{\psfig{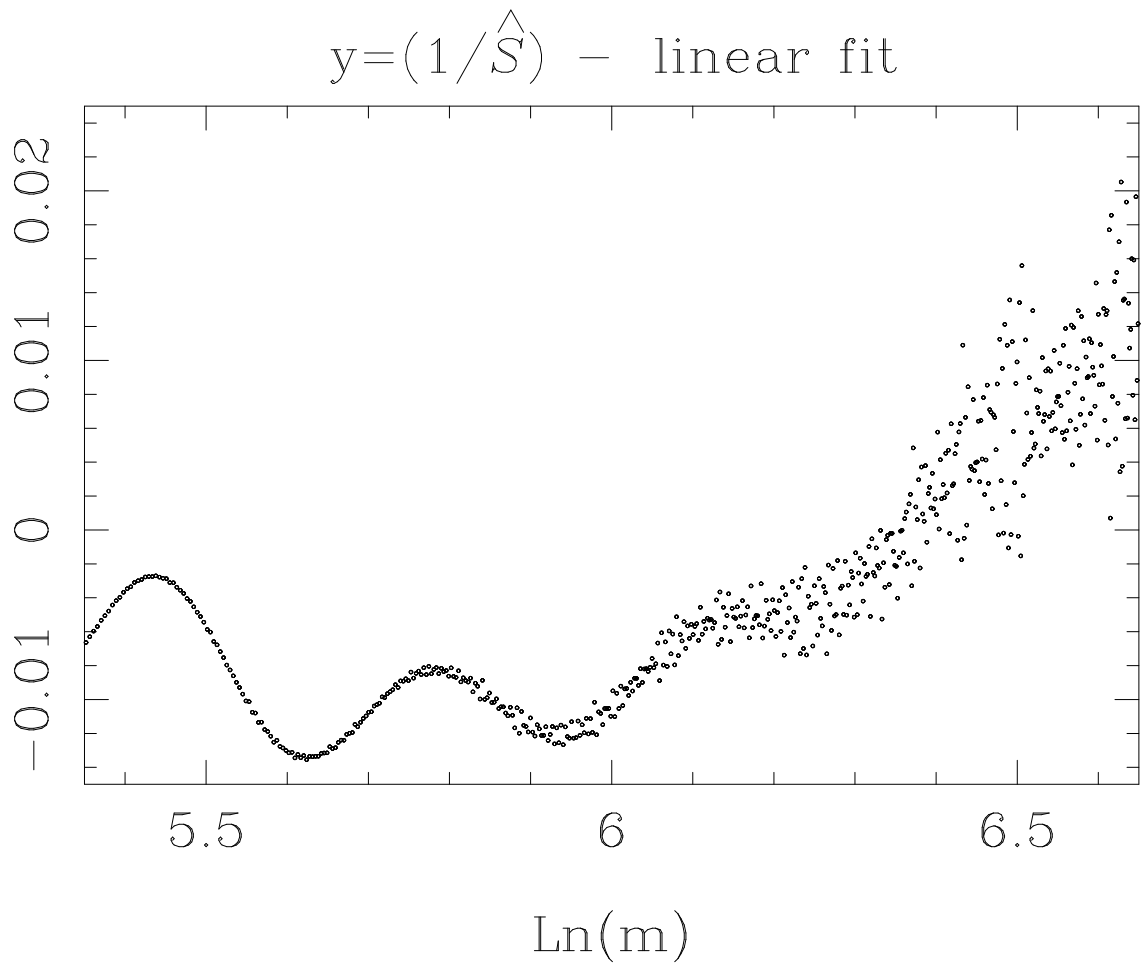}}
\vskip50pt
\caption{Difference between $\widehat{S}_m^{-1}$ and a linear fit 
in $Ln(m)$.}
\label{dinvshat}
\end{figure}

The situation is very different in $D=4$, as shown in Fig. \ref{shat}.
The oscillations are barely visible for low values of $m$, 
and not visible at 
all for larger $m$, where $\widehat{S}_m$ appears to decay smoothly.
If the parametrization of Eq.~(\ref{eq:convpar}) and its
corollary Eq.~(\ref{eq:convdecay}) applied, one might conclude
that $\gamma$ is close to 1.05. However, if we plot 
the inverse of $(\widehat{S}_m)^{-1}$ versus $Ln(m)$, we find the linear
behavior shown in Fig.~\ref{invshat}. This shows that $\widehat{S}_m$
decays like $1/Ln(m)$, so Eq.~(\ref{eq:convdecay}) does not provide
an adequate description of the situation.
The deviation from the linear behavior shows an interesting fine structure
shown in Fig. \ref{dinvshat}. For $m$ near 400 ($Ln(m)$ near 6), ones sees
that the amplitude of oscillation is almost four orders of magnitude smaller
than $(\widehat{S}_m)^{-1}$ itself. For such a values of $m$, the numerical
errors become comparable with the oscillations. For larger values of $m$, 
the numerical errors become larger and wash out the oscillations.
The numerical errors on $(\widehat{S}_m)^{-1}$ in $D=4$
are of the same order of magnitude as what we would estimate 
in $D=3$ from the error
analysis of Ref. \cite{osc} . The main difference is that the oscillations
have a much smaller amplitude in $D=4$. In the following, we will treat the 
oscillations on the same footing as the numerical errors, which
is justified for $m$ sufficiently large.

We will now revisit the asymptotic analysis of $\widehat{R}_m$
and $\widehat{S}_m$
in a more 
general case than Eq. (\ref{eq:convpar}) with $A_0$ and $A_1$ constant.
Our main assumption will be that
\begin{equation}
\chi(\beta)=(1-{\beta \over \beta _c})^{-\gamma }
G(1-{\beta \over \beta _c})\ ,
\end{equation}
where $G$ is such that 
\begin{equation}
lim_{m\rightarrow \infty}
{{G'({\gamma \over m})}\over {mG({\gamma \over m})}}=0
\label{eq:asu}\ . \end{equation}
This restriction includes the case where $G(1-{\beta \over \beta _c})$ 
grows like a positive power of a logarithm
when $\beta$ goes to $\beta _c$. We then proceed as in ref. \cite{gaunt}
and explain the principle of the asymptotic expansion.
We use the residue theorem in the complex $\beta $ plane 
to get an integral representation of 
the coefficients. Next we treat the integral with 
the steepest descent
method. 
Using an exponential parametrization for the integrand of 
the $m$th coefficient,
one finds that the phase has a maximum for a value of $y={\beta\over
{\beta_c}}$ such that 
\begin{equation}
y({\gamma\over{1-y}}-{G'(1-y)\over{G(1-y)}})=m+1 \ .
\end{equation}
The basic principle of calculation is that the second term of the 
l.\ h.\ s.\ of this equation can be treated as a perturbation, for 
$m$ sufficiently large. Neglecting this second term, we get 
\begin{equation}
y=1-{\gamma \over m}+O({1\over{m^2}}) \ .
\label{eq:sadd} \end{equation}
Eq.\ (\ref{eq:asu}) is then seen as the condition which allows us to treat
the second term 
of the 
l.\ h.\ s.\ of Eq.\ (\ref{eq:sadd})
as a perturbation. Finally, 
one finds\cite{gaunt} that 
for large $m$, the leading contribution to the $m$th coefficient 
has the form
\begin{equation}
b_m \propto \beta _c ^{-m} m^{\gamma -1}G({\gamma \over m})
\label{eq:asy}\ . \end{equation}

\begin{figure}
\vskip20pt
\centerline{\psfig{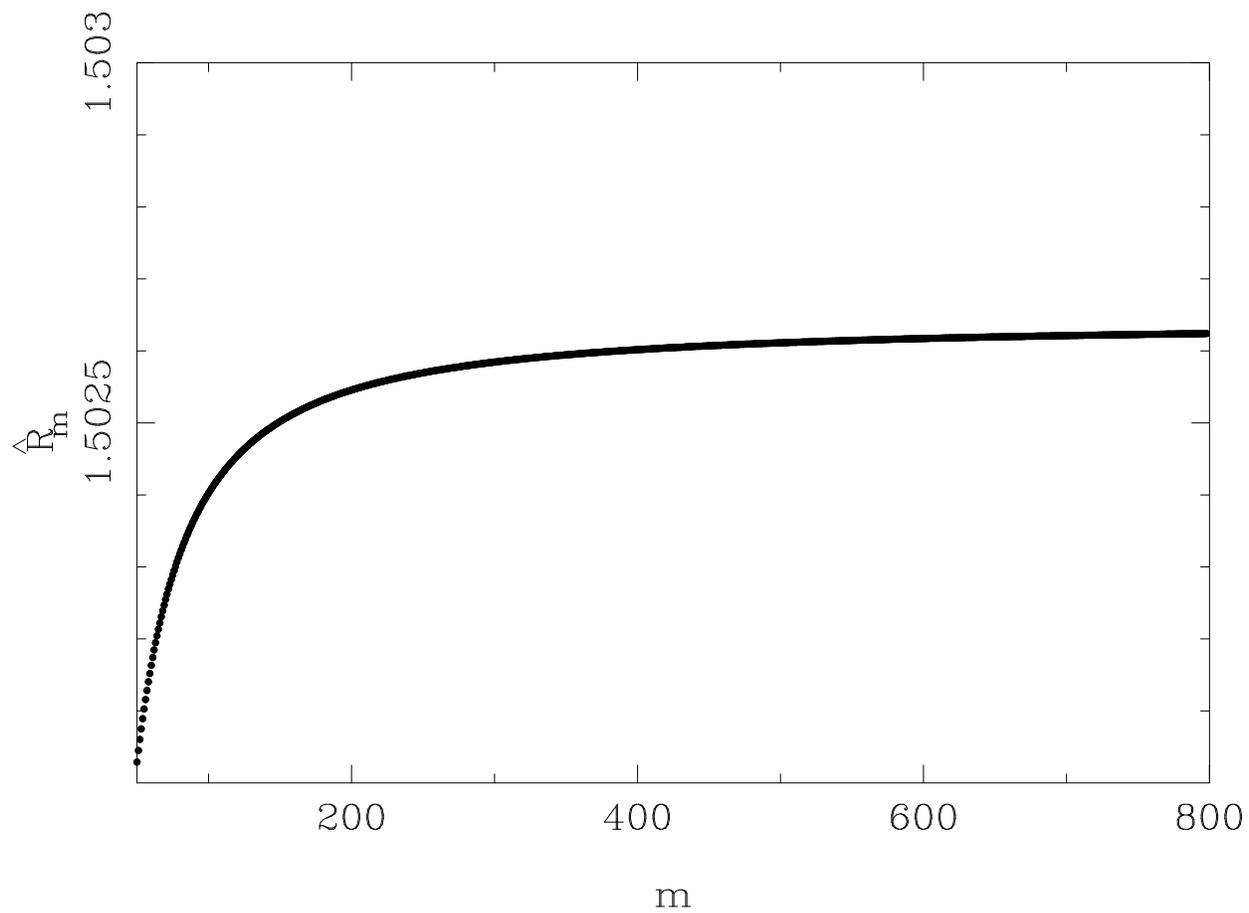}}
\vskip50pt
\caption{
The extrapolated ratio $\widehat{R}_m$ versus the order $m$ in the HT
expansion.}
\label{rhat}
\end{figure}

Before going further, we introduce a parametrization of the ratio
of successive coefficients:
\begin{equation}
r_m=({1\over \beta_c})(1+{1\over m}h(m))
\label{eq:defh}\ . \end{equation}
This definition is independent of any kind of expansion.
From Eq.~\ref{eq:asy}, we found the asymptotic estimate
\begin{equation}
h(m)=(\gamma-1-{\gamma\over m}
{{G'({\gamma \over m})}\over {G({\gamma \over m})}}) +...
\label{eq:hasy} \end{equation}
If we consider the case
\begin{equation}
G(x)\propto (-Ln(x))^p
\label{eq:loghyp}\ , \end{equation}
we obtain
\begin{equation}
\widehat{R}_m=({1\over \beta_c})(1+O({1\over{m (Ln(m))^2}}))\ ,
\end{equation}
and 
\begin{equation}
\widehat{S}_m=\gamma-1+O({1\over{Ln(m)}})\ .
\end{equation}
From this, we can conclude that under the assumption
of Eq.\ (\ref{eq:loghyp}), asymptotic analysis justifies using 
$\widehat{R}_m$  as an estimator for ${1\over \beta_c}$, with estimated
errors on the order of $10^{-4}$. This quantity 
is displayed in Fig.~\ref{rhat}.
As expected, no oscillations are visible. The change between $m=200$ and
$m=800$ is less than $10^{-4}$, which is consistent with 
${1\over{m (Ln(m))^2}}$
corrections. If we use $\widehat{R}_{800}$ as our best estimate, we obtain
$\beta_c=0.665548$, which is in good agreement with our accurate
calculation\cite{finite}, where we found $\beta_c=0.6654955715318593$.
The discrepancy has the same order of magnitude as the small variations
noted above.

On the other hand, for $\widehat{S}_m$,
the corrections to $\gamma -1$ are not very small. For instance,
for $m=800$, $(Ln(m))^{-1}\simeq 0.15$, and it seems implausible that
one could establish that $|\gamma -1|<10^{-3}$ on the grounds of an
expansion in  this not-very small parameter.
More generally, it takes exponentially large $m$ for the ``corrections''
in $(Ln(m))^{-1}$ to become smaller than the ``leading'' $\gamma -1$ when
this quantity is small. Thus it seems desirable to use non-asymptotic 
methods, the subject of the next section.

\section{A Direct Estimation of the critical exponents}
\label{sec:estgam}

In this section, we propose to 
use direct calculations of $h(m)$, defined as
\begin{equation}
h(m)=(r_m \beta_c -1)m\ .
\label{eq:didefh}\end{equation}
This quantity can be 
calculated exactly under some assumption regarding the 
susceptibility, and calculated exactly from the 
empirical series. We emphasize that $h(m)$ is {\it defined} from 
Eq.\ (\ref{eq:defh}) and its
calculation does not require any kind of expansion. However, we need 
to provide an estimate of $\beta _c$. In the following, we will take the 
most accurate value\cite{finite} of 
$\beta_c$ 
quoted in the previous section
rather than the approximate values obtained from $\widehat{R}_m$.

\begin{figure}
\vskip20pt
\centerline{\psfig{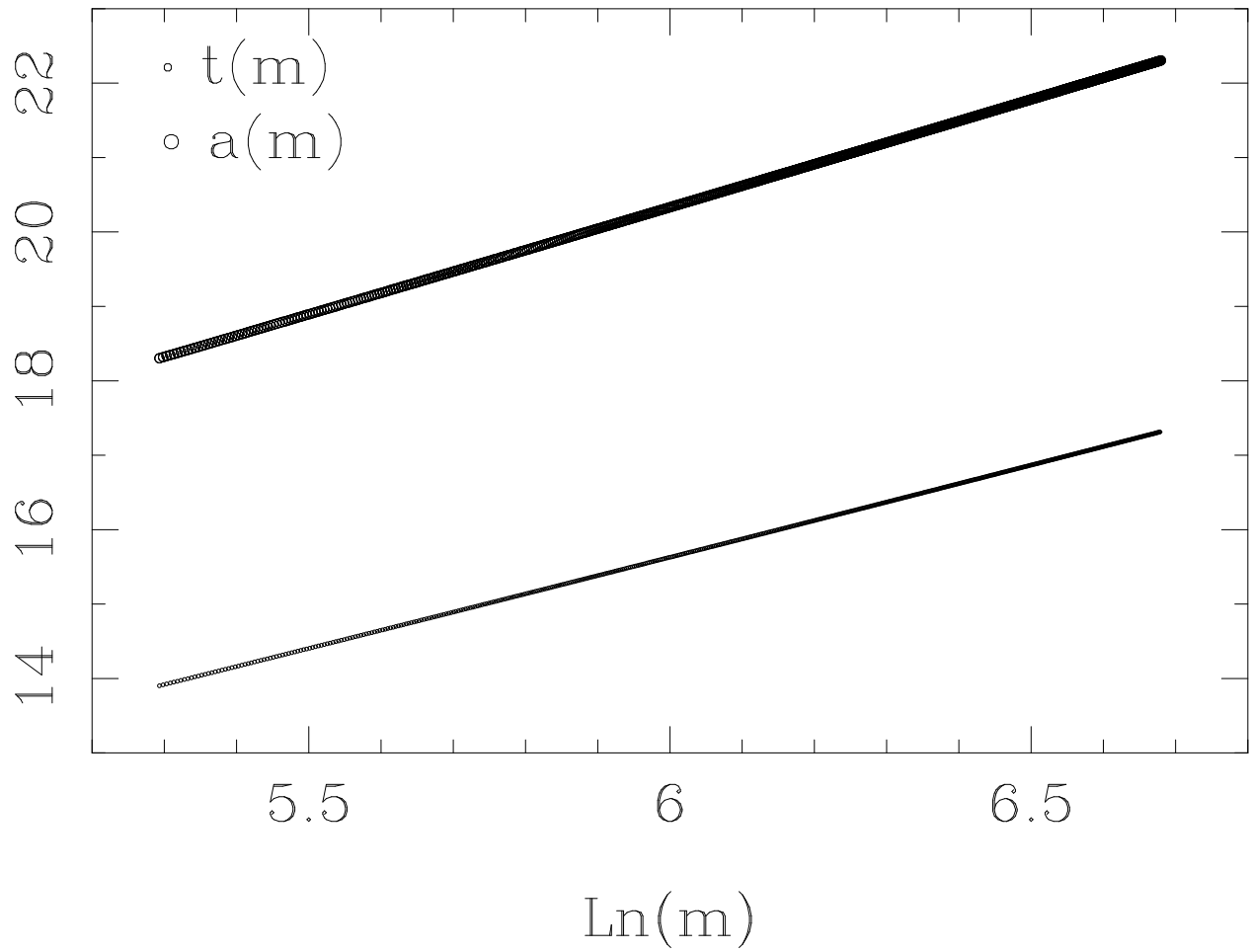}}
\vskip50pt
\caption{$(h(m))^{-1}$ versus $Ln(m)$ from Eq. (\ref{eq:fasone}) 
(large circles, 
above)
and from the actual HT series (small circles, below).}
\label{gap}
\end{figure}

A simple assumption on the leading singularity
of the susceptibility in $D=4$ is given by the result
of a field-theoretical calculation\cite{brezin} 
\begin{equation}
\chi=(\beta _c -\beta )^{-1} (A_0 (|ln(\beta _c -\beta)|)^{1\over 3})\ .
\label{eq:ft} \end{equation}
This lowest-order result would also be obtained for Dyson's model, because
at this order, numerical factors (integrations over the angles),
which are model-dependent, cancel. 
Given that $r_m$ is the ratio 
of two successive coefficients, it is independent of $A_0$ and it
transform as 
$r_m \rightarrow r_m s^{-1}$  under a rescaling $\beta\rightarrow s\beta$.
Consequently, $r_m \beta_c$ is independent of the choice of $A_0$ and 
$\beta_c$. 
We have thus calculated $h(m)$ from the expansion
of 
\begin{equation}
f(x)=(1-x)^{-1}(-{ ln(1-x)\over x})^{1\over 3}
\label{eq:fasone} \end{equation}
in $x$, about $x=0$. The variable $x$ stands for $\beta /\beta _c$.
The division
by $x$ does not change the leading singularity\cite{sykes} 
when $x\rightarrow 1$
while providing a regular expansion around $x=0$.
Under the assumption of Eq.(\ref{eq:ft}), 
we find from Eq.(\ref{eq:hasy}) that asymptotically $h(m)$ 
tends to a small and possibly zero constant plus a correction which decays
like $1\over Ln(m)$. It is thus natural to 
plot $(h(m))^{-1}$ versus $Ln(m)$.
Such a plot is provided in Fig.~\ref{gap}, where 
we compare with $(h(m))^{-1}$
calculated directly from the $D=4$ HT series, using the definition 
Eq. (\ref{eq:didefh}).
The two (approximate) 
lines are separated by an almost constant gap. We tried to modify
the assumption Eq. (\ref{eq:fasone}) in such way that 
the two lines coincide.
The only satisfactory solution we found was the modified assumption
\begin{equation}
f(x)=(1-x)^{-1}((-{ ln(1-x)\over x})^{1\over 3}+r)
\label{eq:fastwo}\ , \end{equation}
where $r$ has to be determined by an error-minimization procedure which we 
now proceed to explain.

\begin{figure}
\vskip20pt
\centerline{\psfig{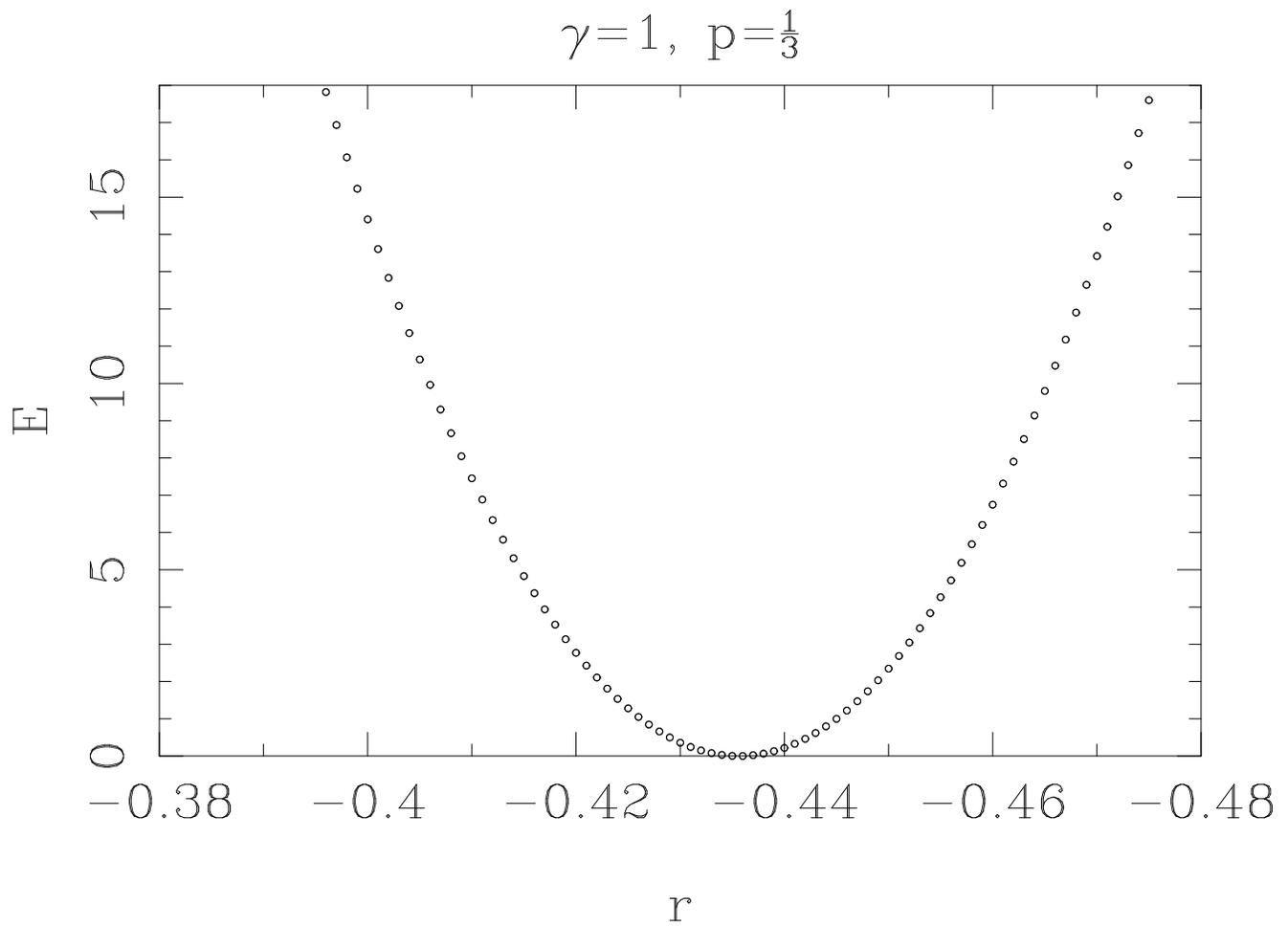}}
\vskip50pt
\caption{The function $E$ defined in Eq. (\ref{eq:err}) versus $r$,
with $a(m)$ calculated from Eq. (\ref{eq:fastwo})}
\label{err}
\end{figure}

\begin{figure}
\vskip20pt
\centerline{\psfig{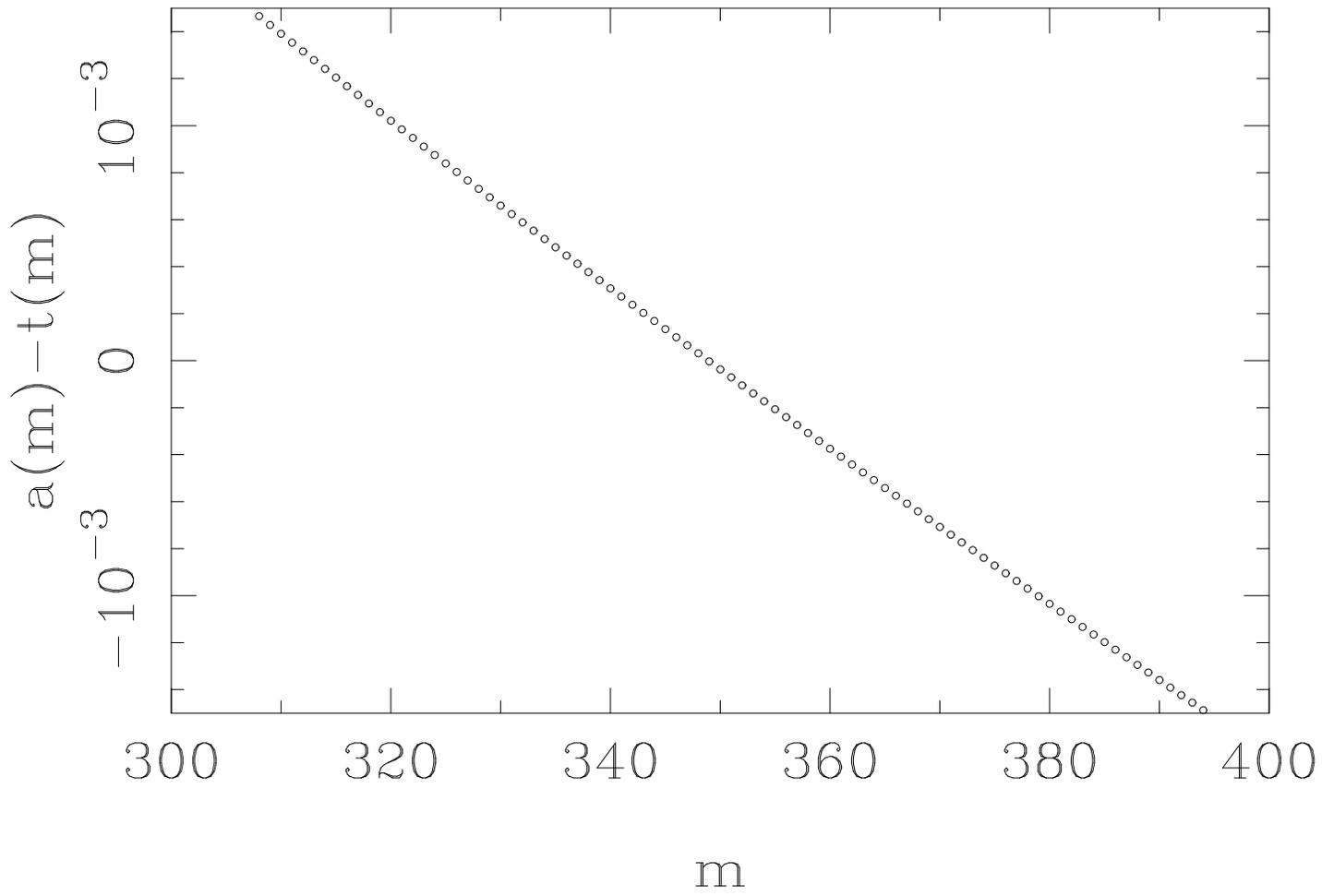}}
\vskip50pt
\caption{$a(m)-t(m)$ versus $m$ for $a(m)$ calculated with
 $r=-0.435622$ in Eq.(\ref{eq:fastwo})}
\label{diff}
\end{figure}

For notational purposes, we call $t(m)$ the ``true'' value of $(h(m))^{-1}$
obtained from the HT series and $a(m)$ the value of $(h(m))^{-1}$
corresponding to an assumed series such as the one obtained from 
Eq.\ (\ref{eq:fastwo}). In practical calculations, we have used the
instruction
$Series$ in $Mathematica$ to calculate $a(m)$. 
It should be noted, that for large orders, rational values of the exponents
give better numerical results. In addition, if the denominator of this
rational exponent gets too large (typically $10^7$ for a calculation up to
order 400), 
one runs out of memory. 
This procedure is quite time-consuming
when one goes to large order. Since such a calculation will 
have to be repeated many times in the rest of this
section, we have used the region $300\leq m\leq 400$ 
to evaluate the discrepancy between $a(m)$ and $t(m)$. As we can
see from the discussion of section ~\ref{sec:lim}, in this range
the oscillations are already small and the numerical errors not too large
yet (see Fig.~\ref{dinvshat}).
We have thus determined the parameter $r$ in Eq. (\ref{eq:fastwo}) by
minimizing
\begin{equation}
E=\sum_{m=300}^{m=400}(t(m)-a(m))^2
\label{eq:err}\ . \end{equation}
The values of $E$ for values of $r$ separated by 0.001 are shown in 
Fig. \ref{err}.
The curve can be fitted very well by a parabola. The minimum of this
parabola is then determined analytically from the 
three values defining the fitting parabola. 
This allows us to find the value
of $r$ with a precision of $10^{-6}$, three orders 
of magnitude smaller than the original resolution. The value 
of $r$ minimizing $E$ found with this procedure is -0.435622, 
corresponding to
a value of $E$ of order $10^{-4}$. Subsequently, we checked this answer by
repeating the calculation of $E$ with a resolution $10^{-6}$ in $r$ and
found the same answer. For this value of $r$, $t(m)$ and $a(m)$
cannot be distinguished in a graph like Fig. \ref{gap}. The difference
between $a(m)$ and $t(m)$ is shown in Fig. \ref{diff}. In the region
where the fit was performed 
($300\leq m\leq 400$), the differences are 4 orders of magnitude 
smaller than the values themselves. 

We conclude from this analysis that Eq.\ (\ref{eq:fastwo}) is a very good
guess concerning the leading and subleading singularities of the 
susceptibility. However, we would like to see if it remains the best guess
if we allow the exponents to change. In other words,
we would to see if different values of the exponents 
could also be acceptable from the point of view of
the high-temperature expansion. We have thus considered a more general
assumption:
\begin{equation}
f(x)=(1-x)^{-\gamma}((-{ ln(1-x)\over x})^{p}+r)
\label{eq:fasgen}\, \end{equation}
and studied $E$ as a function of $\gamma$, $p$, and  $r$. 

\begin{figure}
\vskip20pt
\centerline{\psfig{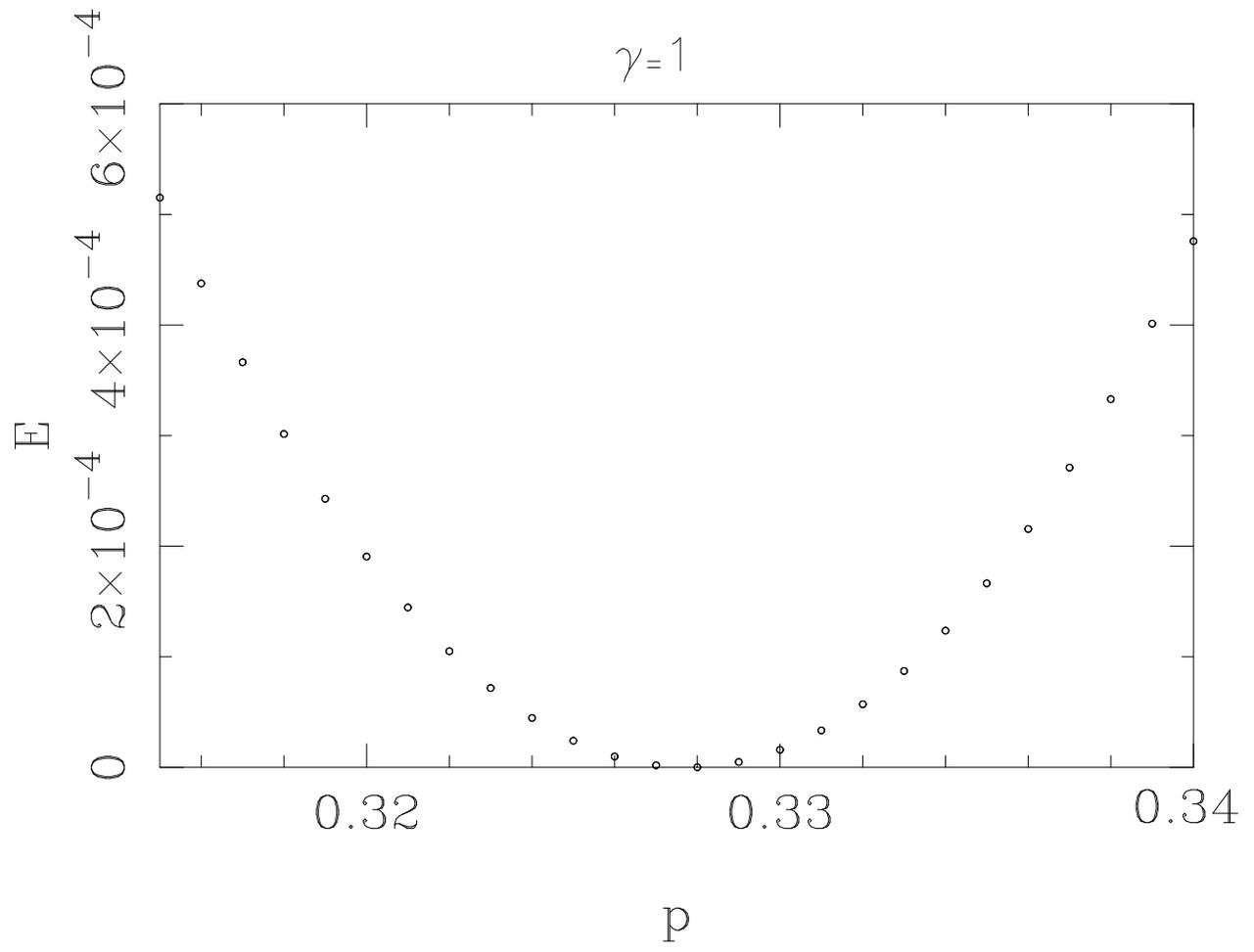}}
\vskip50pt
\caption{The function $E$ versus $p$ with $\gamma =1$ and $r$ chosen 
to minimize $E$.}
\label{erp}
\end{figure}

\begin{figure}
\vskip20pt
\centerline{\psfig{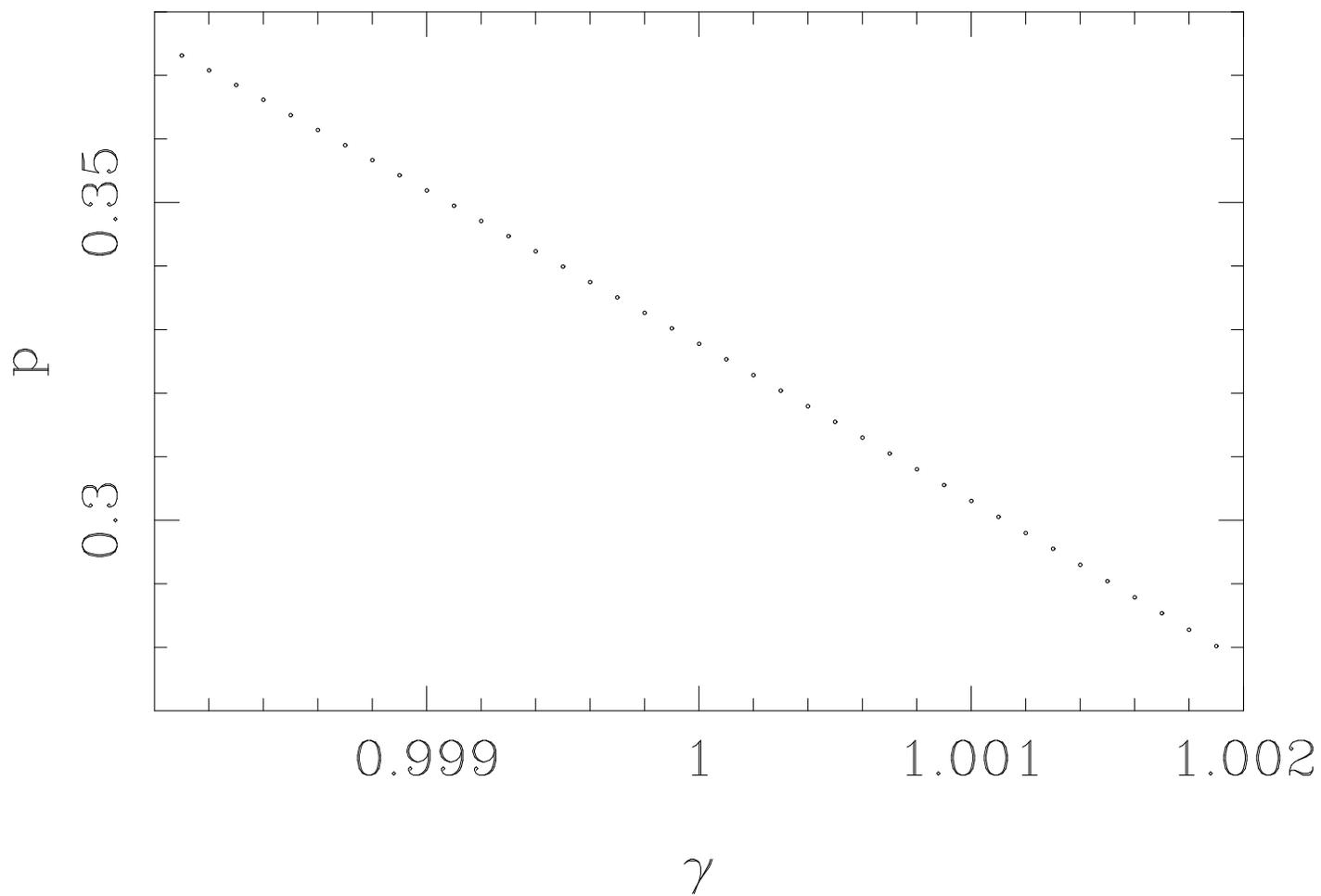}}
\vskip50pt
\caption{Values of $p$ minimizing $E$ at given $\gamma$ and optimum $r$.}
\label{pg}
\end{figure}

\begin{figure}
\vskip20pt
\centerline{\psfig{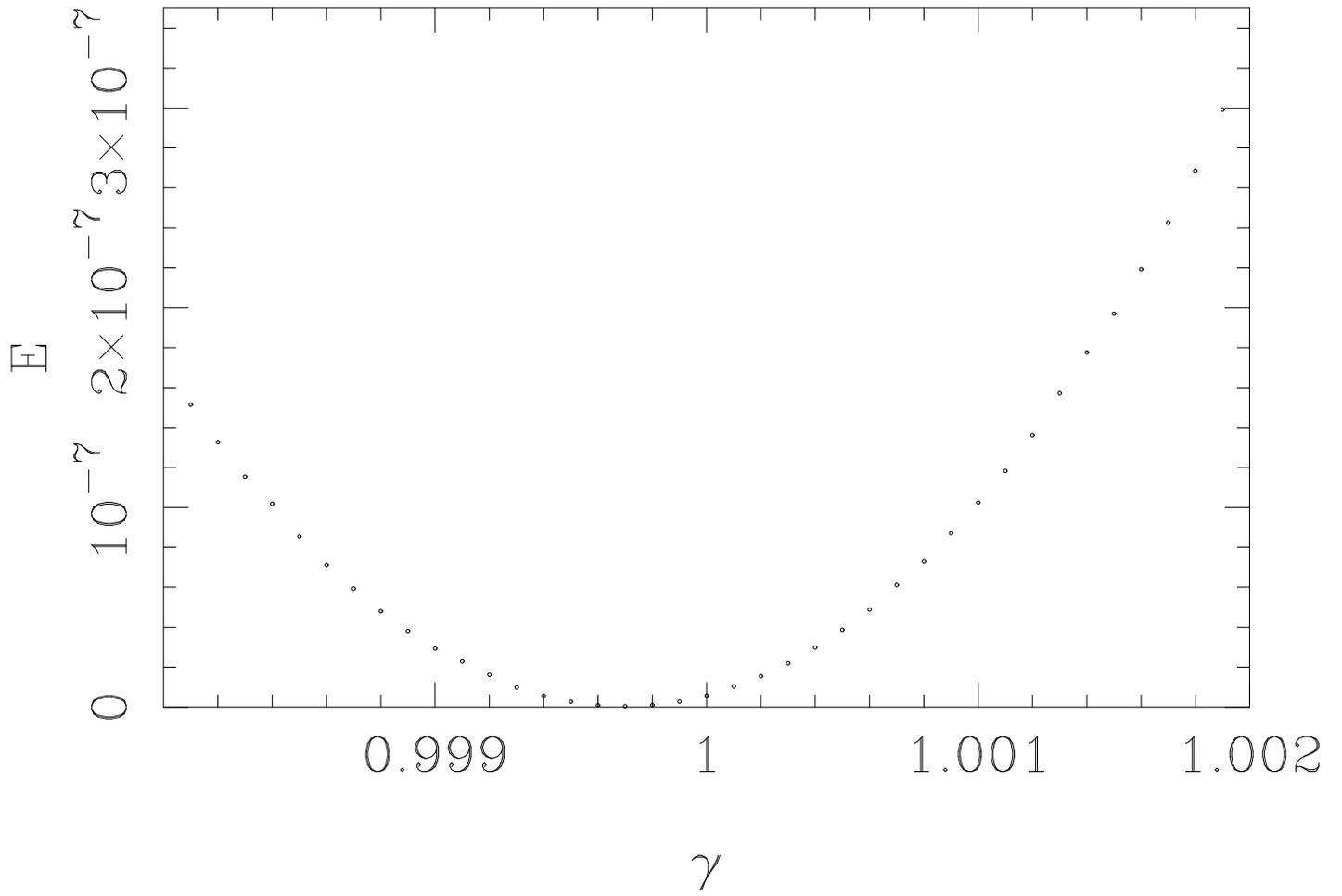}}
\vskip50pt
\caption{The function $E$ versus $\gamma$ with values of
$p$ and $r$ chosen to minimize $E$.}
\label{erg}
\end{figure}

Near a minimum,
$E$ behaves generically as a three-dimensional paraboloid. In this region,
one can ``eliminate'' $r$ by fixing its value in such a way that $E$ is 
minimized with $\gamma$ and $p$ kept constant. The variable $r$ is thus
replaced by a linear combination of $\gamma$ and $p$ plus a constant  
and we can then 
work with a two-dimensional paraboloid. A section of this 
paraboloid defined by the condition $\gamma=1$ is shown in Fig. \ref{erp}.
In a second step, one can 
similarly eliminate $p$ with $\gamma$ fixed by requiring that it takes
the $\gamma$-dependent value that minimize $E$. In the case $\gamma=1$
illustrated in Fig. \ref{erp}, 
this value of $p$ is 0.32775,
not far from the expected\cite{brezin} value $1\over 3$.
This show incidentally 
that a {\it biased} estimate of $p$ is in good agreement with
the field-theoretical result.
Taking values of $\gamma$ separated by $10^{-4}$, we have similarly
calculated the value of $p$ given by the minimization condition.
The results are shown in Fig. \ref{pg}. The linear behavior was
expected: since near the minimum $E$ is a quadratic form, the minimization
condition is linear.
Using this linear relation to eliminate $p$, 
$E(\gamma)$ becomes a parabola. The minimum value taken by 
this function is then the minimum of the initial function $E(\gamma,p,r)$.
This function is shown in Fig. \ref{erg}. $E$ is minimized for
$\gamma=0.9997$, which according to Fig. \ref{pg} corresponds to a value
of $p$ of 0.3351.

In practical calculations, it is convenient to replace parabolic fits
by successive applications of Newton's method. This method has an
adjustable resolution and it allows one to start in regions away from the 
minimum, and where the parabolic behavior does not necessarily hold.

It is difficult to estimate the errors on our result. Since the 
parabolas shown above are reasonably smooth, it seems unlikely that the 
numerical errors or the oscillations, which should have about the same 
size, play any significant role. Most likely, the main source of error
is that $r$ has been considered as a constant.
If instead we allow $r$ to be 
a slowly varying function of $\beta$, we expect in a model independent way,
that these slow variations in $\beta$ will induce slow variations 
in $m$ of the quantity $(h(m))^{-1}$. In
the interval of $m$ considered for the calculation of $E$, the slow
variations can be approximated with polynomials.
In order to get an idea about how low $E$ could become under
such a circumstances, we have fitted the differences 
between $t(m)$ and $a(m)$
displayed in Fig. \ref{diff} and calculated the value of $E$ obtained 
after subtracting these fits from the original differences. 
For a linear fit,
we obtained $E=6\times 10^{-8}$ and for a quadratic 
fit $E=8\times 10^{-10}$.
This shows that by keeping $\gamma=1$ and $p=1/3$ and allowing $r$ to
to be a slowly changing function written in terms of parameters which
are adjusted to minimize $E$,
we can obtain values of $E$ comparable to those 
obtained by keeping $r$ as a constant and allowing $\gamma$, $p$ and $r$ 
to be adjusted in order to minimize $E$.
For definiteness, with $\gamma=1$ and $p$ and $r$ varied to minimize $E$,
we obtain $E=6\times 10^{-9}$. Varying $\gamma,\ p$ 
and $r$, we obtain $E=4\times 10^{-10}$. In conclusion, if we want a more
precise estimation of the critical exponents, we also need more information
regarding the $\beta $-dependence of the subleading singularities.

\section{Triviality and Hyperscaling}
\label{sec:hyper}

Another quantity which can be studied using the high-temperature expansion
is the dimensionless renormalized coupling constant\cite{parisi} 
\begin{equation}
\lambda _4 =- G_4^c m_R^{D+4} \ ,
\label{eq:lam} \end{equation}
where $G_4^c$ is the 
the zero-momentum 
connected Green's function and $m_R$ the renormalized mass.
For $D<4$, this quantity
is designed to have a finite limit when 
$\beta \rightarrow \beta_c$. 
In the case $D=3$, we have checked\cite{finite} 
by a direct calcualtion that $\lambda_4$ reaches the 
value 1.92786 when $\beta \rightarrow \beta_c$. 
In the case $D=4$, we have checked with
a good accuracy that, in the same limit,
$\lambda_4$ goes to zero
like $(Ln(\beta_c -\beta ))^{-1}$.
Thus we have direct evidence that in these two cases
the power singularities cancel in Eq. (\ref{eq:lam}) --- in other
words, that hyperscaling holds.

Bearing in mind that there is no wave function renormalization ($\eta=0$)
in the hierarchical model, we will define as in Ref. \cite{finite} that
$\lambda_4$
is the limit where $n\rightarrow \infty$ of 
\begin{equation}
\lambda_{4,n}= {{{<M_n^4>_n-3(<M_n^2>_n)^2}}
\over{{2^n}({{<M_n^2>_n} \over {2^n}})^{{D\over2}+2}}} \ ,
\end{equation}
with the same notation as in Eq. (\ref{gf}).
Equivalently, with the rescaling factor fixed to $s=\sqrt{2}$ in 
Eq. (\ref{eq:rec})
and the convention of Eq. (\ref{eq:rn}), 
\begin{equation}
\lambda_{4,n}=12{{a_{n,1}^2 -2 a_{n,2}}
\over{(-2a_{n,1})^{{D\over2}+2}}}2^n\ .
\end{equation}

\begin{figure}
\vskip20pt
\centerline{\psfig{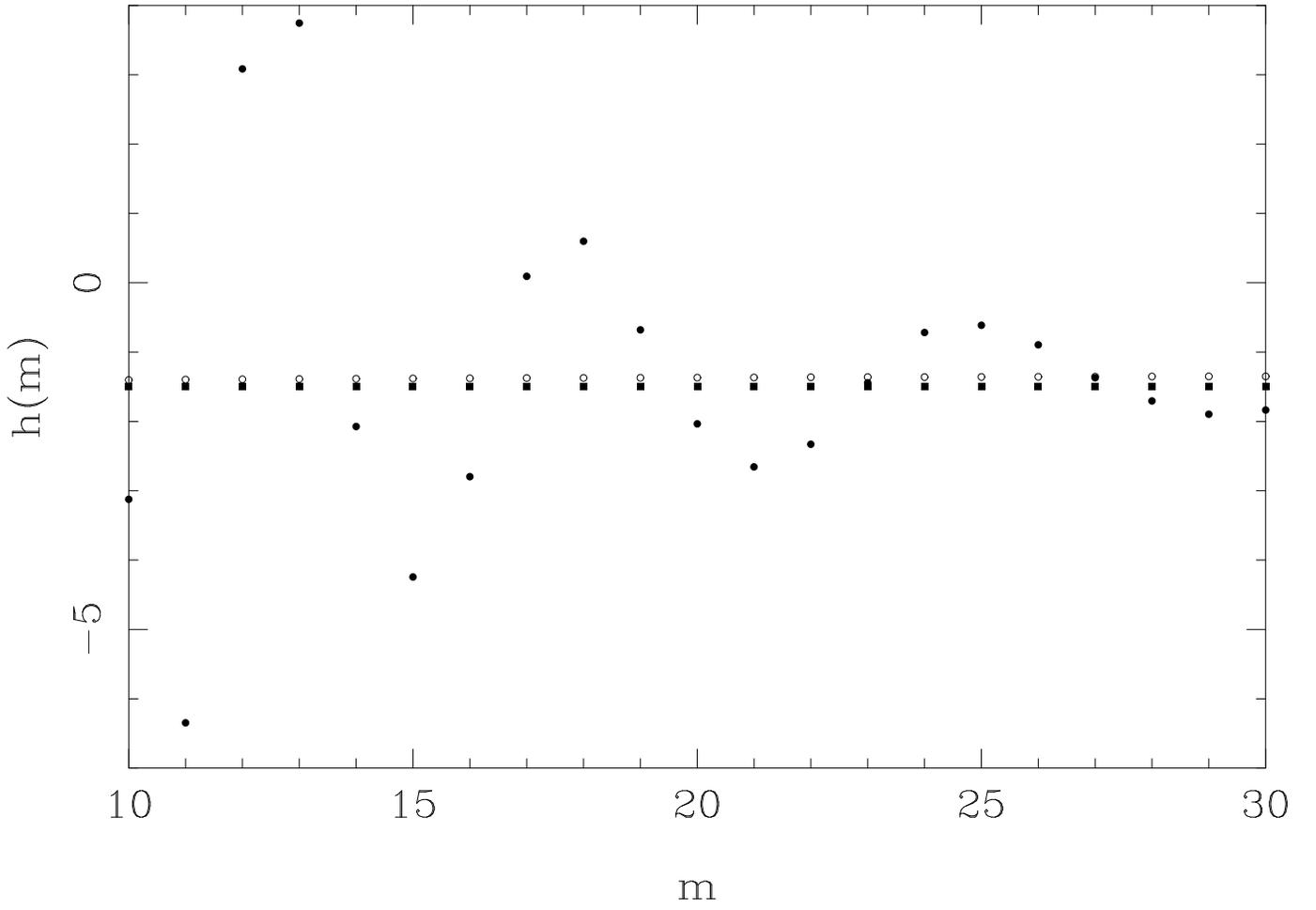}}
\vskip50pt
\caption{The function $h(m)$ corresponding to the HT expansion of $lambda_4$
(dots) compared to the same function for  $-x/Ln(1-x)$ (circles) and 
$(1-x)^{1\over 2}$  (squares). }
\label{hlam}
\end{figure}

The calculation of the HT coefficients of $\lambda_4$ involves the 
subtraction of the disconnected part and it suffers the same type of
numerical problems as the direct calculation of $\lambda_4$, as discussed in
Ref. \cite{finite}. For this reason, we were only able to extract a 
series of 30 coefficients. 
The quantity $h(m)$ defined in Eq.\ (\ref{eq:didefh}) 
corresponding to this series is displayed in 
Fig. \ref{hlam}.
The figure indicates that this quantity has damped oscillations. The average
value of $h(m)$ in the displayed interval is -1.4. From Eq. (\ref{eq:hasy}),
this is consistent with the fact that $\lambda_4$ has a finite limit when 
$\beta \rightarrow \beta_c$, but 
it not possible to distinguish a $(Ln(\beta_c -\beta ))^{-1}$ approach 
to zero from a $(\beta_c -\beta )^{1/2}$ approach.
For comparison, we have displayed in Fig. \ref{hlam}
the function $h(m)$ corresponding to the series generated by  
$-x/Ln(1-x)$  and 
$(1-x)^{1\over 2}$, $x$ being short for ${\beta\over{\beta_c}}$.  
It is clear that the oscillations make the discrimination between these
two asumptions impossible.

\begin{figure}
\vskip50pt
\centerline{\psfig{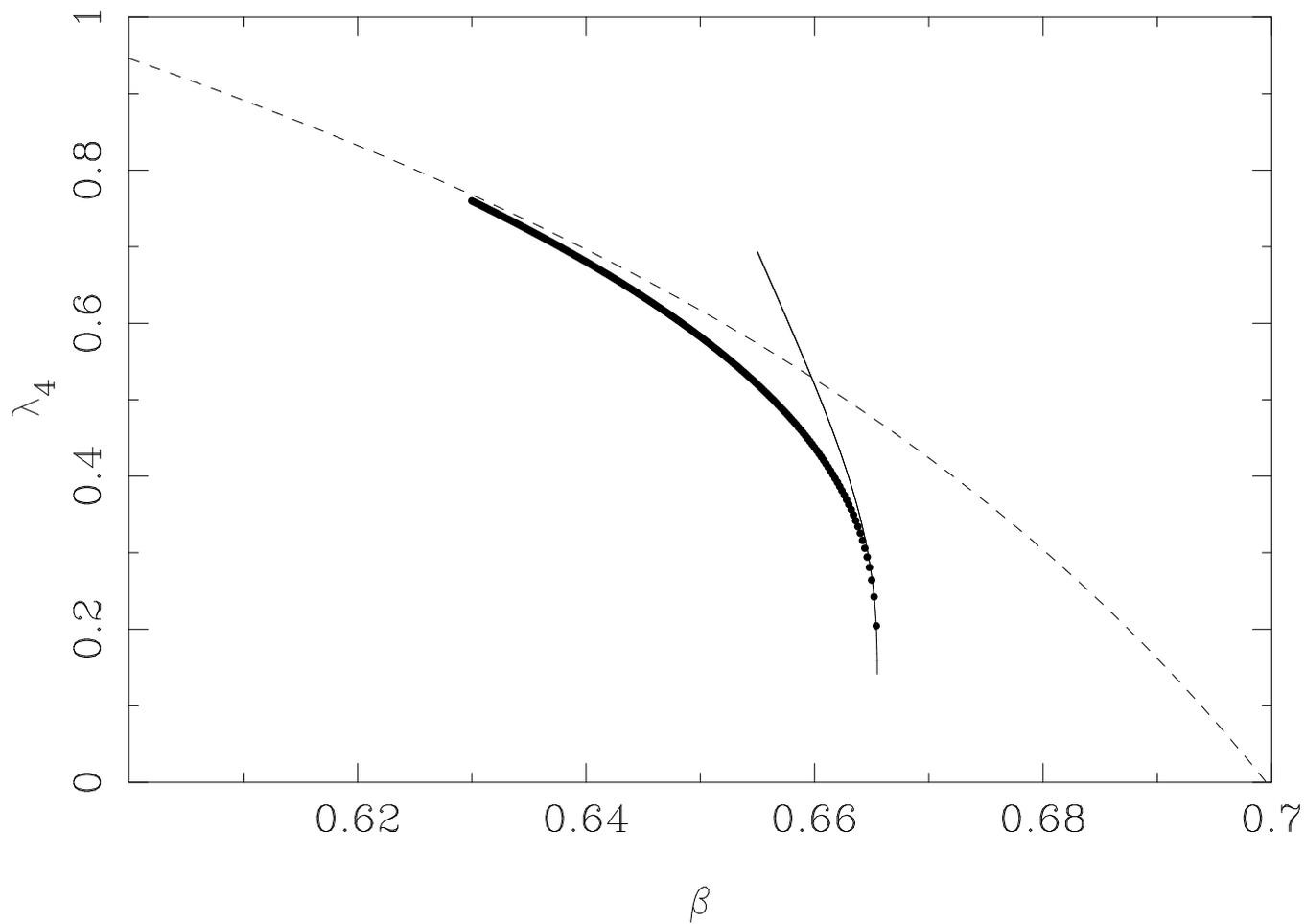}}
\vskip50pt
\caption{$\lambda_4$ versus $\beta$, exact (dots), with the HT 
expansion up to
order 30 (dashed) and in leading singularity (line).}
\label{hyper}
\end{figure}

Another way of seeing that the series is too short to describe the details
of the behavior near $\beta_c$ is to plot the 
truncated expansion of $\lambda_4$ up to order 30. This is done in Fig.\  
\ref{hyper}.
The HT expansion indicates correctly 
that $\lambda_4$ goes to zero when $\beta $
increases. However, the behavior near $\beta_c$ is not accurate.
For comparison, Fig. \ref{hyper} also 
shows the leading critical behavior estimated in Ref. \cite{finite}, namely 
\begin{equation}
\lambda_4 \simeq {1\over{-1.955-0.746\times Ln(\beta_c-\beta)}} \ .
\label{eq:lamcrit}\end{equation}
The data interpolates nicely between the two types of behavior, but we 
see that there is no region in the figure where they overlap.
The order 30 HT expansion gives accurate results for $\beta_c - \beta >
3\times 10^{-2}$, while Eq. (\ref{eq:lamcrit}) becomes accurate when 
$\beta_c - \beta <
10^{-3}$. 

In summary, the truncated expansion makes clear that $\lambda_4$ goes to
zero when $\beta \rightarrow \beta_c$.  In other 
words, the theory is trivial.
However, the series is too short
to extract accurately the precise way it approaches zero, and one cannot
decide from this information whether or not hyperscaling holds.

\section{Conclusions}

There have been questions\cite{dis} 
in the past regarding possible discrepancies between 
field-theoretical calculations based on the renormalization group
approach and calculations based on the high-temperature expansion.
Using a scalar model in the upper 
critical dimension, where all the conventional expansions can be 
compared with direct calculations, we claim that the field-theoretical 
result concerning the leading singularity of the two-point function
at zero momentum given in Eq.\ (\ref{eq:ft}) 
can be reproduced very well
by the high-temperature expansion. 

Using a parametrization of the subleading singularities depending 
on a single
constant $r$, we obtained an optimal agreement 
for the choice $\gamma=0.9997$ and $p=0.3351$. With
this choice, the error on $(h(m))^{-1}$ defined in section \ref{sec:estgam},
is less than one part in a million for $300\leq m \leq 400$. The small
discrepancies between our estimate of the critical exponents
and the field-theoretical values $\gamma=1$ and $p=1/3$ are not 
significant because it is possible to show that small changes in the 
exponents and allowing $r$ to slowly vary have comparable effects for
the quality of the fit.

The present study shows that  
the use of asymptotic analysis or the use of a short series can
be misleading. Given the length of the series available,
asymptotic analysis may be useful for order of magnitude estimates but
not for an accurate determination of the exponents.

There is still 
room for improvement. One could use calculations at fixed $\beta$
to study the corrections to the parametrization of Eq. (\ref{eq:fasgen}).
This procedure could be pursued up to the point where the main source 
of errors would be the numerical errors on the coefficient.

The use of the high-temperature expansion
allows us to probe global features of the renormalization group flows
which cannot be approached using renormalized perturbation theory or an 
analysis of the linearized behavior near the fixed point. 
In particular, our analysis makes implausible, 
for the model considered here, unconventional possibilities
such as the existence in the upper critical
dimension of a non-trivial fixed point
characterized by non-trivial exponents.

\acknowledgments

This research was supported in part by the Department of Energy
under Contract No. FG02-91ER40664. J.J. Godina is supported by
a fellowship from CONACYT.

\end{document}